# Bacterial cell death: Atomistic molecular dynamics simulations reveal a mode of action of structurally nano engineered star peptide polymers

Amal Jayawardena[1], Andrew Hung[2], Greg Qiao[3], Elnaz Hajizadeh[1*]

[1] Department of Mechanical Engineering, Faculty of Engineering and Information Technology, University of Melbourne, Parkville, VIC, 3010, Australia
[2] School of Science, STEM College, RMIT University, VIC, 3001, Australia
[3] Department of Chemical Engineering, Faculty of Engineering and Information Technology, University of Melbourne, Parkville, VIC, 3010, Australia
*Corresponding Author: ellie.hajizadeh@unimelb.edu.au

## Abstract

Multidrug resistance (MDR) to conventional antibiotics is one of the most urgent global health threats, necessitating the development of effective and biocompatible antimicrobial agents that are less inclined to provoke resistance. Structurally Nanoengineered Antimicrobial Peptide Polymers (SNAPPs) are a novel and promising class of such alternatives. These star-shaped polymers are made of a dendritic core with multiple arms made of co-peptides with varying amino acid sequences. Through a comprehensive set of *in vivo* experiments, we (*Nature Microbiology,* **1**, *16162, 2016*) showed that SNAPPs with arms made of random blocks of lysine (K) and valine (V) residues exhibit sub-$\mu M$ efficacy against Gram-negative and Gram-positive bacteria tested. Cryo-TEM images suggested pore formation by SNAPP with random block co-peptide arms as one of their mode of actions. However, the molecular mechanisms responsible for this mode of action of SNAPP were not fully understood. To address this gap, we employed atomistic molecular dynamics simulation technique to investigate the influence of three different sequences of amino acids, namely 1) alternating block KKV 2) random block and 3) di-block motifs on secondary structure of their arms and SNAPP's overall configuration as well as their interactions with lipid bilayer. We, for the first time identified a step-by-step mechanism through which alternating block and random SNAPPs interact with lipid bilayer and leads to 'pore formation', hence cell death. These insights provide a strong foundation for further optimization of the chemical structure of SNAPPs for maximum performance against MDR bacteria, therefore offering a promising avenue for addressing antibiotic resistance and development of effective antibacterial agents.

## 1. Introduction

Multidrug resistant (MDR) bacteria are a growing global health crisis, posing a significant threat to both public health and economy[1-4]. MDR superbugs contribute to an estimated 700,000 deaths annually, and without interventions, this mortality rate is anticipated to escalate to 10 million deaths per year by 2050[5-7]. The demand for alternative antibacterial treatments that are less inclined to provoke antimicrobial resistance has never been more pressing.

Antimicrobial peptides (AMPs) offer a promising solution in the relentless battle against MDR bacteria[8-10]. These remarkable agents, distinct from conventional antibiotics, employ a multifaceted approach to combat bacterial cells, including disrupting and permeabilizing membranes, forming pores, and ultimately causing cell demise[9,11]. Within this array of tactics, AMPs sculpt pores within the bacterial inner membrane through barrel-stave and toroidal pore mechanisms, while also employing non-pore forming strategies such as carpet model, where peptides cover the bacterial membrane and disrupt its shape[11,12]. While membrane disruption remains central, AMPs also exhibit a range of other mechanisms, as some are known to traverse membranes to target internal components[11,13]. Factors such as type of amino acid residues, charge, hydrophobicity, and overall structure significantly affect their antimicrobial activity, with secondary structure playing a pivotal role[10,14].

The secondary structures of AMPs are controlled by amino acid arrangement and charge[10,15-21]. The formation of secondary structures such as α-helices or β-sheets is governed by noncovalent interactions among residues within the peptide sequence[10]. Consequently, altering a specific sequence can disrupt these interactions, potentially leading to a different secondary structure[16-19]. AMPs encompassing α-helical and β-sheet structures are both recognized for their involvement in membrane disruption, albeit with varying specificities[10,15]. Generally, a higher tendency to form α-helices results in more potent antimicrobial activity, where their activity is correlated with their amphipathicity[10,21,22]. The role of helicity and its connection to AMP antibacterial activity has been extensively investigated[23-26]. While the degree of helicity holds importance, equal consideration must be given to the environmental context in which the AMP assumes this structure and its secondary structure flexibility[27-29].

While AMPs hold great promise in combating MDR bacteria, they are not without their challenges. These challenges include cytotoxicity, susceptibility to protease degradation, short *in vivo* half-life and manufacturing complexity[11,30]. These obstacles have led researchers to explore alternative innovative solutions, one of which is Structurally Nanoengineered Antimicrobial Peptide Polymers (SNAPPs)[31-33]. SNAPP represents a novel category of antibacterial peptides introduced by Qiao *et al.* in 2016[31]. SNAPP possesses peptide arms extending outward from a nano-engineered core, resulting in an overall star-like architecture. In contrast to existing self-assembled antimicrobial macromolecules, which break down into individual units below their critical micelle concentration, SNAPPs exhibit remarkable stability as single-molecule structures even at extremely low concentrations. In experimental studies, SNAPP have demonstrated effective ability to disrupt bacterial cell membranes[31]. However, the efficacy of SNAPP is not solely determined by their mere presence but is influenced by several factors, including number of arms, length of arms, and the specific sequence of amino acids within each arm[33]. From circular dichroism (CD) experiments, it has been shown that SNAPP arms adopt an alpha helix secondary structure in a hydrophobic environment[31-33]. The strategic placement of amino acids in an arm emerges as a crucial factor that impacts the flexibility and secondary structure, which in turn plays a significant role in SNAPPs' antimicrobial activity against MDR bacteria[27,28].

In our pursuit of a deeper understanding of SNAPP and its mechanism of action against multidrug-resistant (MDR) bacteria, we recognize the value of molecular dynamics (MD) simulations[34-38]. While multiple molecular dynamics studies have investigated AMPs[39-41], it is worth highlighting the scarcity of literature dedicated to SNAPP. These simulations allow us to unravel the atomic-level details of the stepwise process by which SNAPP disrupt bacterial cells, shedding light on the intricacies of their cell destruction mechanisms. Moreover, MD simulations provide an opportunity to investigate the secondary structure of SNAPP in both hydrophilic and hydrophobic environments, offering insights into how these peptides interact with cells under different conditions[42-44]. It is important to note that previous atomistic MD studies of pore forming AMPs were performed on a pre-defined pore on the bacteria cell membrane. As far as we are aware, this study for the first time shows the step-by-step mechanism of pore formation and contributing chemical structure factors.

In this investigation, we developed three types of model SNAPPs: an alternating block SNAPP, featuring a repeating sequence of lysine (K) and valine (V) (KKVKKVKKVKKVKKV), random block SNAPP, featuring a random sequence of lysine and valine (KKKVVKKKVVKKKVK) and a di-block SNAPP, characterized by two separate blocks of lysine and valine comprising each arm (KKKKKKKKKKVVVVV). This comparative analysis aims to explore the impact of amino acid placement and sequence order on the antimicrobial efficacy of SNAPP. By examining whether amino acid placement affects the antimicrobial effect, we can gain a comprehensive understanding of the structural factors influencing SNAPP activity.

Literature on MD simulation of SNAPP is limited with only one previous study, using coarse-grained model, providing some insights into their interactions with lipid bilayers[45]. However, coarse-grained models, while informative, lack the atomic-level details and hence the ability to accurately predict peptide structural transitions resulting from exposure to different environments, which are necessary to fully elucidate the step-by-step mechanism of SNAPP action. Therefore, our MD simulations seek to bridge this knowledge gap and offer a more comprehensive understanding of SNAPPs' antimicrobial mechanisms through all-atomistic MD simulation of SNAPP interaction with bacterial cell membrane.

This paper is structured as follows. In Section 2, a detailed description of the molecular dynamic simulations is provided. Section 3 details the results and discussion, which is divided into three distinct parts. Section 3.1. focuses on the analysis of environment-dependent secondary structure of arms. Section 3.2. presents results on the radius of gyration of the modelled SNAPPs in polar and non-polar environments, while Section 3.3. delves into investigating the bactericidal mechanism of SNAPP. Finally, in conclusion, a summary of key findings is provided.

## 2. Methodology

In our study, we utilized the GROMACS 2021.V3 molecular dynamics simulation software package to model the interactions between SNAPP and a model gram-negative bacterial lipid bilayer[46]. We employed the CHARMM36 force field, which has been used in numerous studies to elucidate membrane properties and peptide-lipid interactions[47]. All modelled SNAPP chemical structures, i.e., the alternating block SNAPP, random block SNAPP and di-block SNAPP feature eight arms with distinct amino acid sequences as shown in Figure 1c.

To construct the structure of an eight-arm SNAPP, a stepwise approach was followed. Firstly, the individual arms of SNAPP were modelled using the Avogadro molecular modelling software, which allowed us to define the specific amino acid sequences and initial secondary structure of each arm[48]. In order to define the initial secondary structure we utilized the AlphaFold[49] and I-TASSER[50] deep learning structure prediction tools to predict the overall structure of two models of SNAPP arm as shown in Table 1. Based on structural predictions, we constructed the arms with alpha-helices as the initial structures for the SNAPPs. Additionally, AlphaFold2 indicated alpha helix, while I-TASSER indicated a mixture of helix, strand, and coil for the di-block arm. Next, the nanoengineered core of the SNAPP was generated using the Packmol software[51]. The core is approximated as a hydrophobic sphere, with 400 carbon atoms to mimic a non-polar spherical surface. Finally, the individual arms were bonded to the core via the C-terminus carbonyl carbon to a selected core atom using a defined covalent bond with bond length of 2.5 Angstroms, creating the final structure of the eight-arm SNAPP. This step involved aligning and joining the arms with the core in a precise manner, ensuring proper connectivity and maintaining the desired spatial arrangement of the arms around the central core.

Table 1 - Secondary structure prediction results for alternating, random and di-Block SNAPP arms using AlphaFold and I-TASSER prediction tools (H, S, and C Represent Helix, Strand, and Coil)

| SNAPP model | Alpha Fold | I-TASSER |
| --- | --- | --- |
| **Alternating block** KKVKKVKKVKKVKKV | Alpha Helix | Alpha Helix CCHHHHHHHHHHHCC |
| **Random block** KKKVVKKKVVKKKVK | Alpha Helix | Alpha Helix CHHHHHHHHHHHHCC |
| **Di-block** KKKKKKKKKKVVVVV | Alpha Helix | Mix of helix, strand, and coil CCCHHHHHSSSSSSC |

To perform secondary structure and radius of gyration analyses, all three types of SNAPP models were initially placed in a water solution and simulated for 100 nanoseconds. These three SNAPP systems were enclosed in a box with dimensions of 4.984 x 5.410 x 4.441 nanometres, containing 14,665 TIP3P water molecules and 80 chloride ions to neutralize the system. Subsequently, these three SNAPPs were also solvated in a box containing both TFE and water, with the same dimensions of 4.984 x 5.410 x 4.441 nanometres. This box contained 2,666 water molecules, 2,650 trifluoroethanol (TFE) molecules, and 80 chloride ions.

The bilipid membrane atomistic model was constructed using the CHARMM-GUI input generator, specifically utilizing the membrane builder tool within this server[52]. To create the modelled bilipid membrane mimicking the inner membrane of a gram-negative bacteria, a combination of 1-palmitoyl-2-oleoyl-sn-glycero-3-phosphatidylethanolamine (POPE) and 1-palmitoyl-2-oleoyl-sn-glycero-3-phosphatidylglycerol (POPG) lipids with a ratio of 4:1 was generated in the upper and lower leaflets (see Figure 1d).

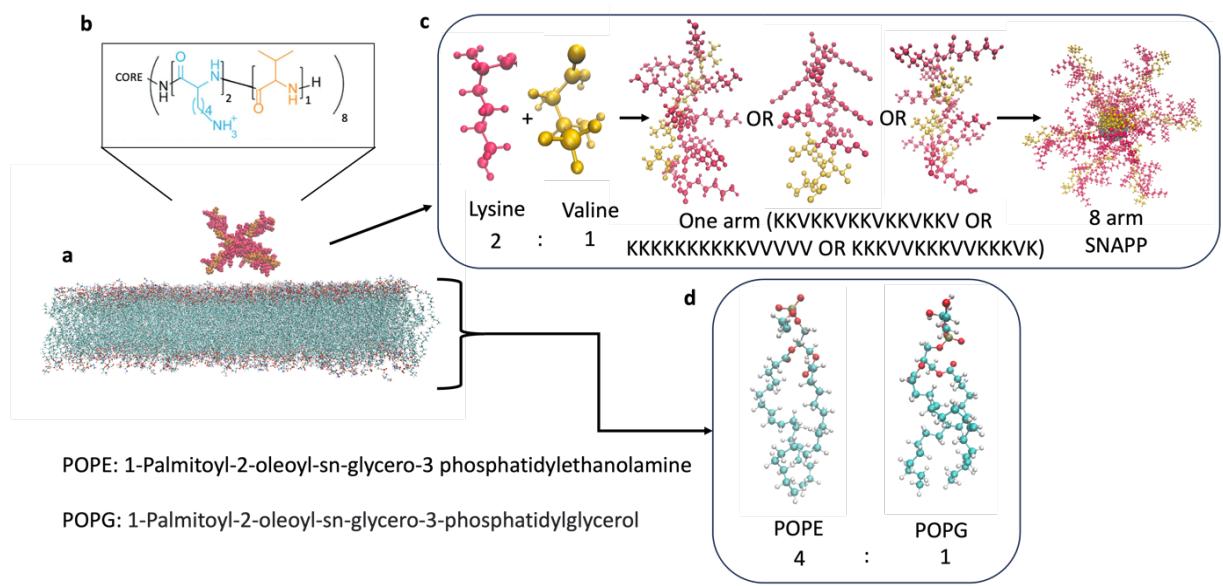

Figure 1: (a) The snapshot of eight-armed SNAPP positioned 2 angstroms above the upper leaflet of the bilipid layer. (b) The chemical structure of the modelled eight-armed SNAPP. (c) The chemical structure of eight-arm SNAPP with lysine and valine residues. Alternating block SNAPP has eight arms of KKVKKVKKVKKVKKV, di-block SNAPP has eight arms of KKKKKKKKKKVVVVV and random block has eight arms of KKKVVKKKVVKKKVK. (d) The chemical structure of the modelled bilayer. The bilipid model consists of POPE and POPG lipids with 4:1

To place the SNAPP structure above the bilipid layer, the Visual Molecular Dynamics (VMD) software was utilized[53]. This facilitated the integration of the SNAPP coordinates with the bilipid layer, ensuring that the SNAPPs were positioned at a distance of 2 Å from the top of the bilipid layer as shown in Figure 1a. Positioning SNAPP 2 Å away from the lipid bilayer's top surface reduces any bias in initial binding orientation by permitting re-orientation upon the start of the simulation. This practice aligns with established protocols in the literature[39,45]. The simulation box for bilipid, measuring 7.96940 x 7.96940 x 7.96940 nanometres, contained 146,674 TIP3P water molecules, 1,080 POPE lipids, and 270 POPG lipids. Subsequently the system was neutralized by introducing $Na^+$ and $Cl^-$ ions to create a neutral system of ~150 mM salt concentration. After that, the energy of the system was minimized using the steepest descent algorithm until the maximum force change was less than 1000 kJ·mol$^{-1}$·nm$^{-1}$. Following this, the system was equilibrated for 750 picoseconds. The LINear Constraint Solver (LINCS) algorithm was used to constrain H-bonds. Particle Mesh Ewald (PME) method was used to do the electrostatic calculations. During NPT equilibration steps, the temperature was set at 303.15 K, using the Berendsen thermostat with a time constant of 1 ps. Pressure was kept at 1 bar using the Berendsen barostat with semi-isotropic pressure coupling at a time constant of 5 pico-seconds and an isothermal compressibility of 4.5e-5 (kJ·mol$^{-1}$·nm$^{-3}$)$^{-1}$. During the production runs, the temperature was kept at 303.15 K using the Nose-Hoover thermostat with a time constant of 1 ps. The pressure of 1 bar was maintained using the semi-isotropic Parinello-Rahman barostat with a time constant of 5 ps. The isothermal compressibility remained at 4.5e-5 (kJ·mol$^{-1}$·nm$^{-3}$)$^{-1}$. The production runs were executed five times, each with different initial random seeds for starting velocities ensuring a statistically representative average. Unless otherwise stated, all data were calculated as averages over the five replicate trajectories. Each run, for the SNAPPs in water and water/TFE solvents, lasted for a duration of 100 nanoseconds. For the simulations investigating the interactions between bilipid and SNAPP, 50 nanosecond simulations were performed. Subsequently, secondary structure analysis and radius of gyration calculations were performed.

## 3. Results and Discussion

### 3.1. Secondary structure analysis of alternating-block, random block and di-block SNAPPs in polar and non-polar environments

To verify the reliability of our modelled SNAPP structures and simulation methodology, we examined the secondary structure of 8-arm alternating block SNAPP model and compared them to previously obtained experimental results[31-33].

As mentioned in the introduction, incorporating findings from CD spectroscopy experiments[54], previous research established that SNAPP arms with random blocks of lysine and valine exhibit an alpha helix secondary structure in a hydrophobic environment and show a random coil secondary structure in water[31-33]. To validate our model, we performed molecular dynamic simulations that placed the modelled alternating block, random block, and di-block SNAPPs in a solvent mimicking a partially hydrophobic environment comprising a mixture of trifluoroethanol (TFE) and water.

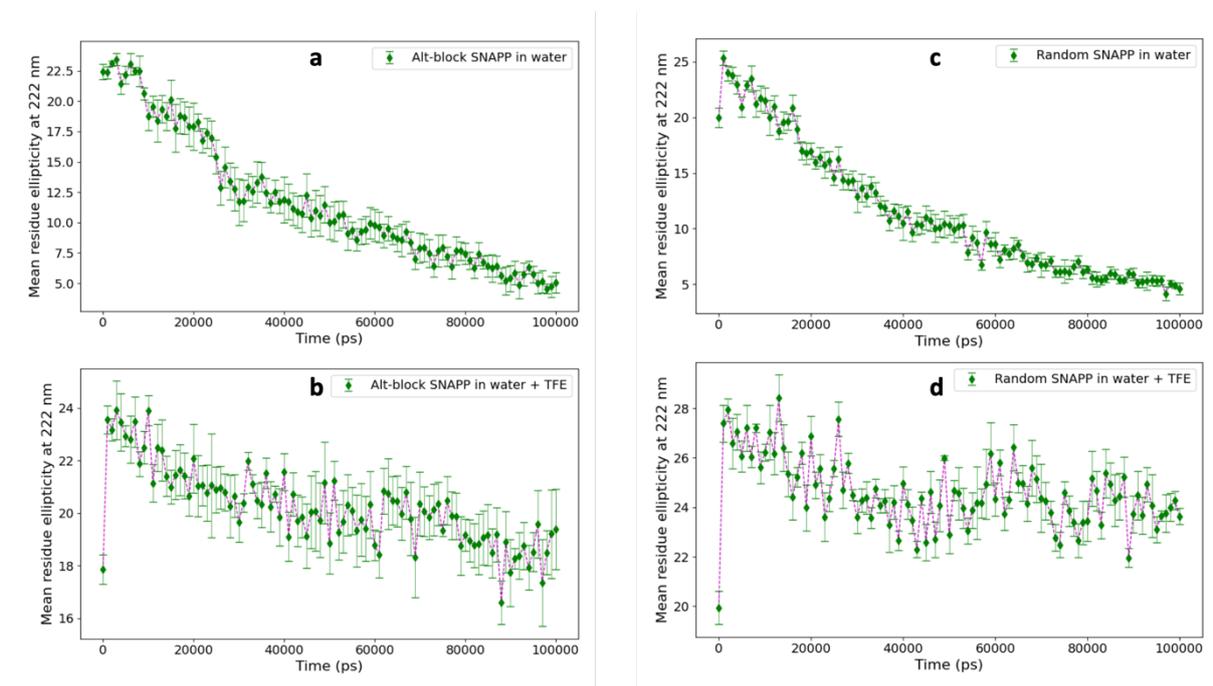

Figure 2: (a) Ellipticity at 222 nm for Alternating block SNAPP with the amino acid sequence KKVKKVKKVKKVKKV in water over time. (b) Ellipticity at 222 nm of alternating block SNAPP in water + 80% v/v (volume ratio) TFE, illustrating stable alpha helix secondary structure in hydrophobic environments. (c) Ellipticity at 222 nm for random block SNAPP with the amino acid sequence KKKVVKKKVVKKKVK in water over time. (d) Ellipticity at 222 nm of random block SNAPP in water + 80% v/v (volume ratio) TFE, illustrating stable alpha helix secondary structure in hydrophobic environments.

Simulation of alternating block SNAPP in water was conducted for 100 ns to investigate its secondary structure. An alpha helix secondary structure was initially assigned to the simulation to validate experimental observations. However, as shown in Figure 2a and Figure S1, the alpha helix structure did not remain stable in water for alternating block SNAPP, consistent with experimental findings indicating that SNAPP exhibits a random coil secondary structure in hydrophilic environments such as water. As depicted in Figure 2a, the ellipticity of the alternating block SNAPP arm, initially ranging from 20-25 nm, significantly decreases to nearly zero by 100 ns, confirming the stable random coil secondary structure of the alternating block SNAPP arm in water.

Subsequently, 100 ns simulations were conducted for alternating block SNAPP in a solution containing water with an 80% v/v (volume ratio) TFE concentration to explore their secondary structure in a hydrophobic environment. Starting with an alpha helix secondary structure, Figure 2b and Figure S2 illustrate the stability of the alpha helix structure throughout the simulation in the water + TFE solution for the arms of alternating block SNAPP. This outcome aligns with experimental results, confirming that the 8-arm alternating block SNAPP forms an alpha helix structure in a hydrophobic environment and reinforcing the consistency of the predictions of our MD model with the experimental observations. Similar observations were seen for the random block SNAPP as shown in the Figures 2c and 2d.

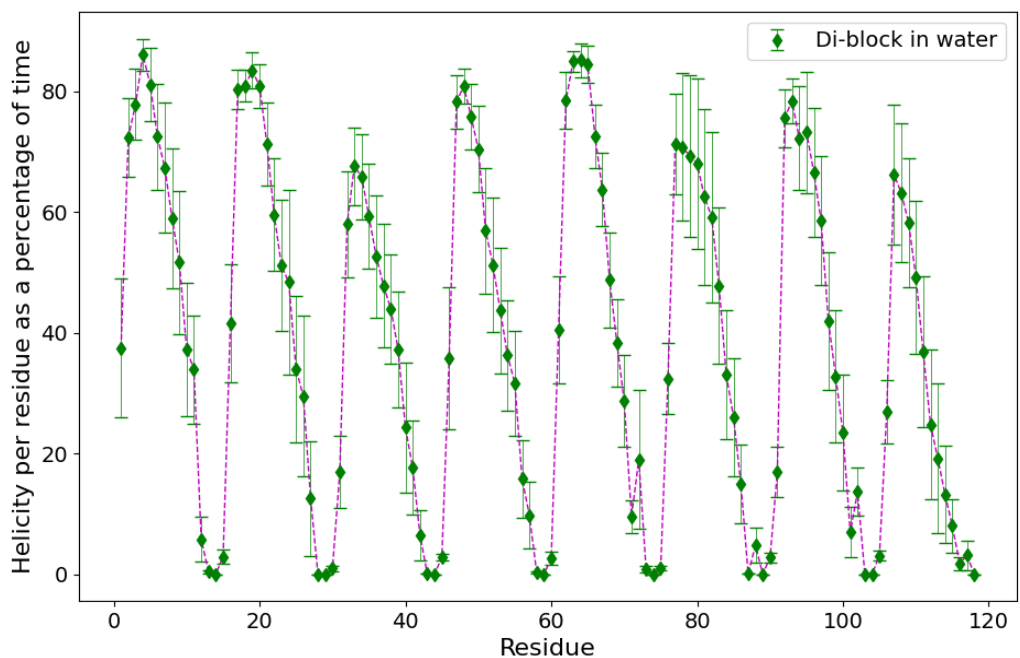

Figure 3: Helicity per residue of the 8-arm di-block SNAPP in water illustrates that 5 out of 8 arms maintain their alpha helix secondary structure for more than 80% of the simulation

As depicted in Figure 3, unlike the alternating and random block SNAPPs, the di-block SNAPP retains its helicity in water over the same period. Consequently, it can be concluded that, while the secondary structure of alternating and random block SNAPPs is highly sensitive to their environment, where they adopt a random coil and an alpha helix in hydrophilic and hydrophobic environments, respectively. In contrast, the di-block SNAPP lacks such a structural flexibility. This analysis contributes to our understanding of the factors influencing the varying cell disruption capabilities observed among different SNAPPs.

## 3.2. Radius of gyration analysis of alternating-block, random block, and di-block SNAPPs in polar and non-polar environments

In addition to conducting secondary structure analysis, we performed radius of gyration analysis in both polar and non-polar environments for the modelled SNAPPs to quantify the extent of size variation exhibited by the three modelled SNAPPs in water and water + TFE environments. An alpha helix secondary structure was initially assigned to all the SNAPPs models.

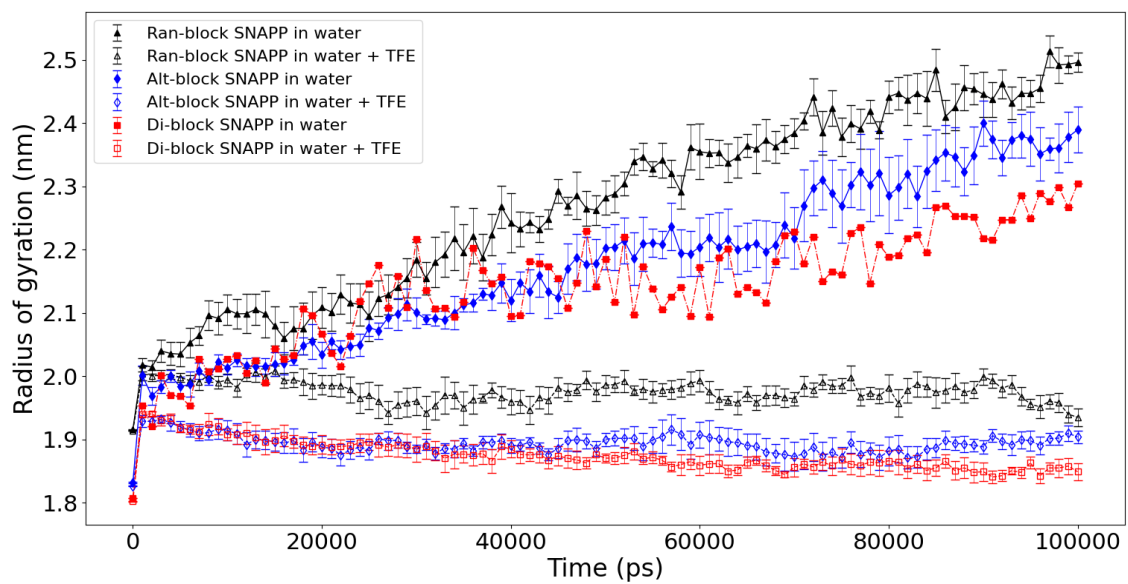

Figure 4: Radius of gyration (nm) analysis was performed for modelled SNAPPs, where diamond markers represent the alternating block SNAPP, square markers represent the di-block SNAPP, and triangular markers represent the random block SNAPP. Filled symbols are data in water and open symbols are data in water + TFE for all systems.

Figure 4 reveals significant findings. Firstly, in water, the random block SNAPP exhibits the highest variation in radius of gyration, ranging from 2.0 nm to 2.5 nm, while it displays the lowest variation in the water + TFE (non-polar) environment. Conversely, the di-block SNAPP demonstrates the lowest radius of gyration variation among the three modelled SNAPPs in water. Notably, in the water + TFE environment, the di-block SNAPP has the highest radius of gyrations variation compared to alternating and random block SNAPP models.

## 3.3. Killing mechanism by SNAPP via pore formation and membrane disruption

In previous laboratory experiments[38-40], it was demonstrated that SNAPP disrupts bacterial cell membranes. Current study employs atomistic molecular dynamics simulations to uncover the precise mechanism by which SNAPP disrupts bacterial cell membranes. Shedding light on this intricate process, not only enhances our understanding of SNAPP's mode of action but also paves the way for fine-tuning its structure for future applications.

To achieve these objectives, molecular dynamics simulations lasting 50 ns were conducted, involving alternating and di-block SNAPPs placed atop the lipid bilayer, as illustrated in Figure 1a. The simulations aimed to accomplish two main goals: firstly, to predict the initial binding mechanism of SNAPP to the membrane structure, and secondly, to elucidate subsequent step-by-step interactions of SNAPP with bilipid, resulting in disruption events which may be critical to its bacterial killing mechanism.

### 3.3.1. Alternating-block SNAPP adopts an 'octopus-like' configuration and binds strongly to bacterial membranes

Figure 6 displays snapshots of a series of interactions taking place between alternating SNAPP molecule and lipid bilayer leading up to bacterial cell membrane disruption. Within the first 10 ns of this MD simulation, the alternating block SNAPP displayed binding to the lipid bilayer while going through a notable configurational transformation. The SNAPP molecules shifted from their original star-shaped configuration in Figure 6a to an expanded arm (octopus-like) structure, as visually represented in Figure 6b. This alteration in SNAPP's configuration which happens around 20 ns, corresponds to an increased formation of hydrogen bonds between its arms and the bilipid layer components, i.e., POPE and POPG as shown in Figure 6c. This observation suggests that SNAPP arms not only anchored themselves within the lipid bilayer but also established robust interactions with the neighbouring lipid molecules. These interactions contribute to their stability and may play a crucial role in advancing the subsequent steps of their kill mechanism. The octopus-like configuration results in a large contact footprint on the membrane surface, maximising the interactions between the SNAPP and bilayer.

During the whole 50 ns simulation of the alternating block SNAPP, one notable observation was its gradual submergence into the lipid bilayer while maintaining their octopus-like configuration structure characterized by widened, spread arms as shown in Figure 6e. This submergence process was visually evident and supported by partial density data, as shown in Figure 8a. Additionally, a tilting motion of the SNAPP towards one side was observed during the submerging phase. This tilting behaviour could be attributed to the stronger hydrophobic attraction of a specific arm's valine residues with the hydrophobic lipid tails as shown in Figure 6d using dashed circle, resulting in SNAPP being pulled further into the lipid bilayer.

Furthermore, the number of hydrogen bonds increased over time as shown in Figure 6c. Notably, during the 50 ns of interaction of SNAPP with the lipid bilayer, the amino acid lysine formed a greater number of hydrogen bonds compared to valine in the arms of the SNAPP. This observation was supported by separate bond analysis conducted for lysine and valine. The graph in Figure 6c clearly demonstrates that lysine formed approximately 25 hydrogen bonds during the first 50 ns, whereas valine formed only approximately 5 hydrogen bonds via their peptide backbone regions.

The membrane insertion depth for SNAPP is investigated through calculating density profiles for both SNAPP and lipid bilayer (Figure 8a), showing that the alternating-block SNAPP spreads out on the lipid surface, forming a single peak. The high degree of overlap in the density plots with the lipids is indicative of a substantial insertion. Furthermore, it shows that all the arms participate in lipid contact (Figure 8c, blue bars) within a short time of simulation, consistent with the wide contact footprint described above.

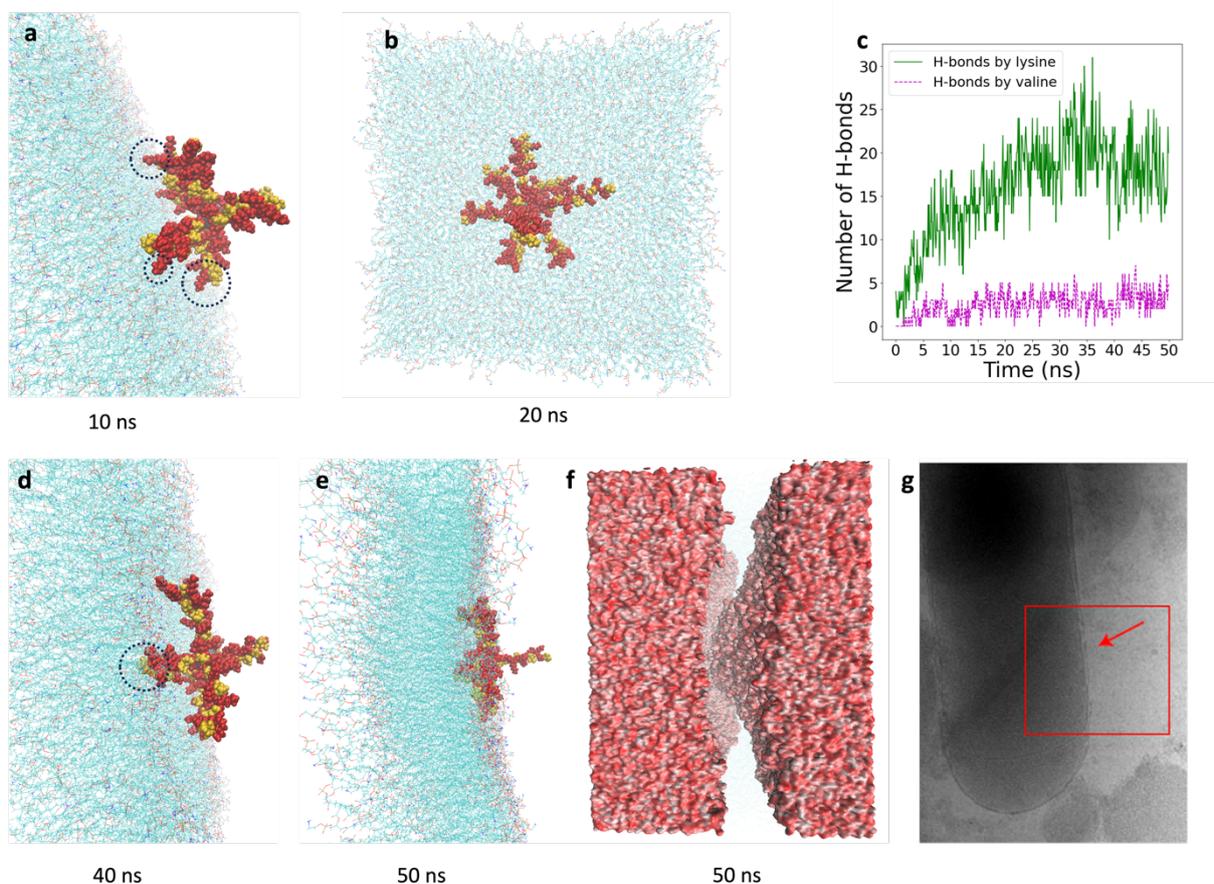

Figure 6: Illustration of the step-by-step mechanism by which alternating block SNAPP disrupts the bilipid membrane. The red colour represents lysine residues, and the yellow colour represents valine residues. (a) Depicts the initial binding of SNAPP to the lipid bilayer via electrostatic attractions highlighted by the dashed circles. (b) Shows the extension of SNAPP's arms, resulting in an octopus-like configuration as it adheres to the lipid bilayer. (c) Graphically represents the increased H-bond formation of lysine compared to valine during the simulation. (d) Highlights the penetration of SNAPP into the bilipid layer, facilitated by hydrophobic valine residues highlighted from the dashed circle. (e) Illustrates SNAPP's immersion into the bilipid layer after 50 ns. (f) Displays the pore formation and disruption of the bilayer due to the entry of water into the bilayer red colour depicts water molecules. (g) Cryo-TEM images of stripped cell walls and membranes of E. coli after the incubation with S16 SNAPP for 90 min at a lethal dose of 35 µg ml$^{-1}$. With permission from "Combating multidrug-resistant Gram-negative bacteria with structurally nanoengineered antimicrobial peptide polymers," by Lam S.J et. al, 2016, Nature Microbiology, 1, 16162.

### 3.2.2 Di-block SNAPP adopts an extended 'pufferfish-like' morphology and binds weakly to bacterial membranes

Figure 7 displays snapshots of MD simulation of di-block SNAPP interacting with lipid bilayer. The di-block SNAPP takes a different configuration compared to that of the alternating-block SNAPP upon contact with the bilayer. Instead of adopting a flat configuration on the surface, arms of the di-block SNAPP remain extended even after making contact with the lipid bilayer. This is likely to be due to the strong repulsion between adjacent arms due to a high concentration of positively charged poly-lysine blocks, resulting in a 'pufferfish-like' configuration, where the peptide arms form extended spikes which make initial contact with the membrane surface.

As a result of its extended configuration, the di-block SNAPP did not adhere to the bilipid membrane in the first 10ns. Visual analysis indicated that the valine amino acid residues at the end of the SNAPP arms failed to establish the necessary hydrogen bonds required for adhesion to the bilipid as depicted in the Figures 7a-b. This lack of adhesion can be attributed to the hydrophobic nature of valine amino acids, which hinders their attraction and ability to penetrate past the lipid headgroup region. Additionally, the high density of positive charges concentrated in the poly-lysine region of the di-block SNAPP vastly reduces the ability of this segment to enter the non-polar core region of the membrane. Thus, both blocks exhibit non-favourable interactions with at least one region of the lipid membranes.

In the subsequent 50 ns simulation of the di-block SNAPP, it became apparent that the di-block SNAPP encountered challenges in adhering to the lipid bilayer. The lysine residues within the arms exhibited visible attempts to form hydrogen bonds with the lipid head groups. As shown in graph 7c it is evident that lysine and valine did not exhibit a similar trend in the formation of hydrogen bonds, unlike what we observed with the alternative block SNAPP. While the arms positioned closer to the bilipid demonstrated a tendency to dock, the arms far from the bilipid grappled with establishing the necessary bonds. This is also evident in the number of contact analysis shown in Figure 8c, where it clearly demonstrates that only the arms closer to the bilipid membrane make contacts with the bilayer. This disparate behaviour in the arms stemmed from the concentration of hydrophobic amino acids on one side of the arm and hydrophilic amino acids on the opposite side of the arm in the di-block SNAPP as shown in Figure 7a-e. Across the five simulations conducted for the di-block SNAPP, two distinct peaks emerged in the partial density as shown in Figure 8b because of this interaction. These peaks indicate a repulsion between the hydrophobic content of certain arms and the hydrophilic content of others within the star-shaped structure. This repulsion impedes the arms from forming bonds with the bilipid, resulting in the maintenance of the star-like structure, as opposed to the widening of arms observed in the alternating block SNAPP. As evident from Figure 8c, almost all eight arms of the alternating block SNAPP contacted the bilipid, while for the di-block SNAPP, only four arms made contact. This leads to the di-block SNAPP's inability to induce membrane disruption as evident from the partial density graph shown in Figure 8b.

The lower membrane insertion depth of di-block SNAPP is evident through their calculated density profile (Figure 8b) showing that SNAPP formed two density peaks, with the left peak corresponding to a few arms which make contacts with the membrane, and the right peak corresponding to the arms which are strongly repelled from the membrane, extending out into the water solvent. The relatively low degree of overlap in the density plots with the lipids is indicative of only shallow insertion. Furthermore, only a smaller number of arms participated in lipid contact (Figure 8c, orange bars) throughout the simulation, consistent with the much smaller footprint and narrow spike-like insertion mechanism of di-block SNAPPs described above.

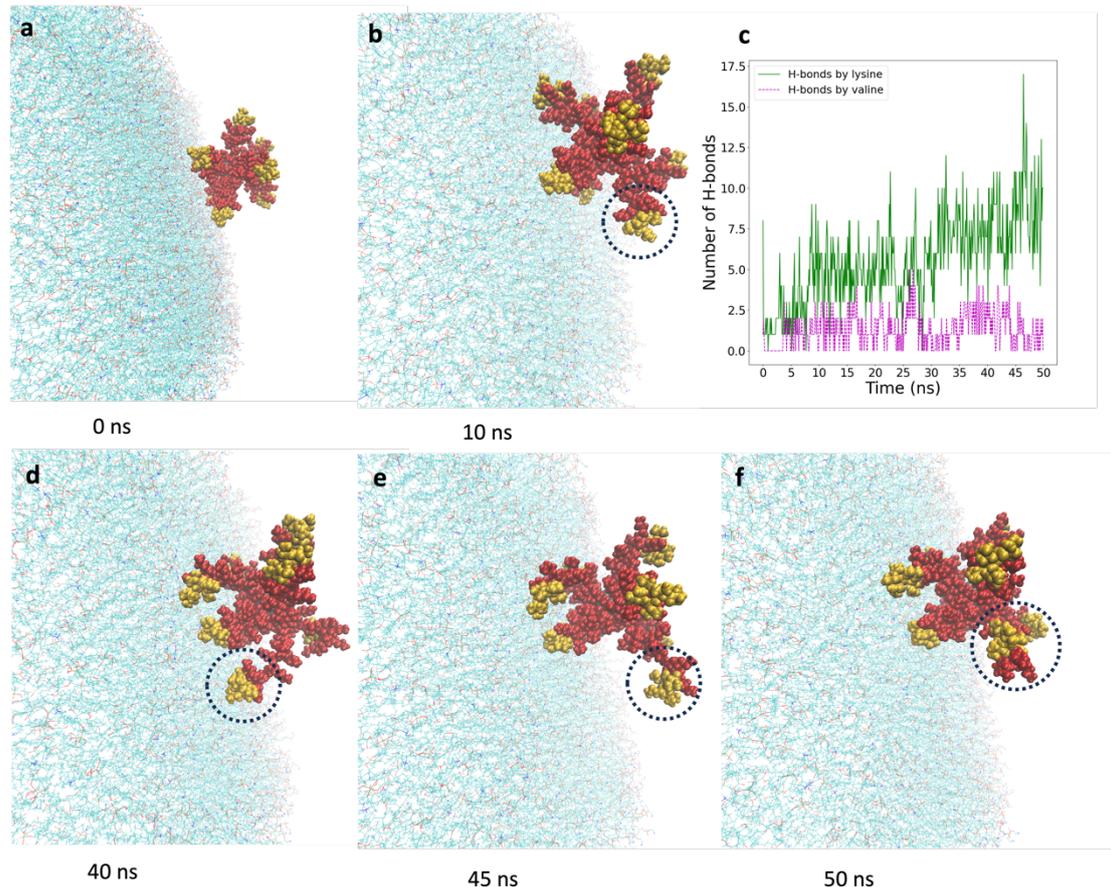

Figure 7: Interaction of di-block SNAPP with the bilipid membrane. The red colour represents lysine residues, and the yellow colour represents valine residues. (a) Illustrates the initial arrangement of the di-block SNAPP. (b) Shows that at the 10 ns mark, the di-block SNAPP's valine residues attempt to adhere to the bilipid membrane, particularly the valine residues marked with a dashed circle. (c) Depicts that at the 20 ns mark, the marked valine residues repel away from the bilipid due to charge densities of the hydrophobic part interacting with the densely charged distribution of the lysine residues. (d)-(f) Depict the same scenario as described above.

Thus, overall, the alternating block SNAPP exhibits a structure whereby its peptide arms extend outwards the membrane plane, maximising its contact footprint on the bilayer ('octopus' mechanism). In contrast, the di-block SNAPP retains an extended configuration, owing to strong electrostatic repulsion among adjacent arms via the highly dense continuous stretches of lysine residues. This configuration results in only a limited number of arms being capable of forming contacts with the membrane (spike-like) and attaching onto the surface via the N-termini of these arms in a much more limited fashion ('pufferfish' mechanism).

Experimental data supports this observation, as evidenced by alternating block SNAPP's lower Minimum Disruptive Concentration (MDC) values compared to di-block SNAPP. For instance, against gram negative *E. coli*, alternating block SNAPP exhibits an MDC of 0.8 µM, whereas di-block SNAPP requires 17.3 µM. Similarly, for the gram-positive S. aureus, alternating block SNAPP demonstrates an MDC of 1 µM, while di-block SNAPP necessitates over 100 µM[31]

Based on the secondary structure analysis, it is evident that the random block SNAPP exhibits similar variations in secondary structure to the alternate block SNAPP. Therefore, it is reasonable to assume that the random block SNAPP will have a similar killing mechanism to the alternate block SNAPP. Moreover, these molecular dynamics simulation findings suggest the importance of secondary structure flexibility in determining SNAPP's antimicrobial activity, highlighting the superior performance of alternating and random block SNAPPs over the di-block sequence.

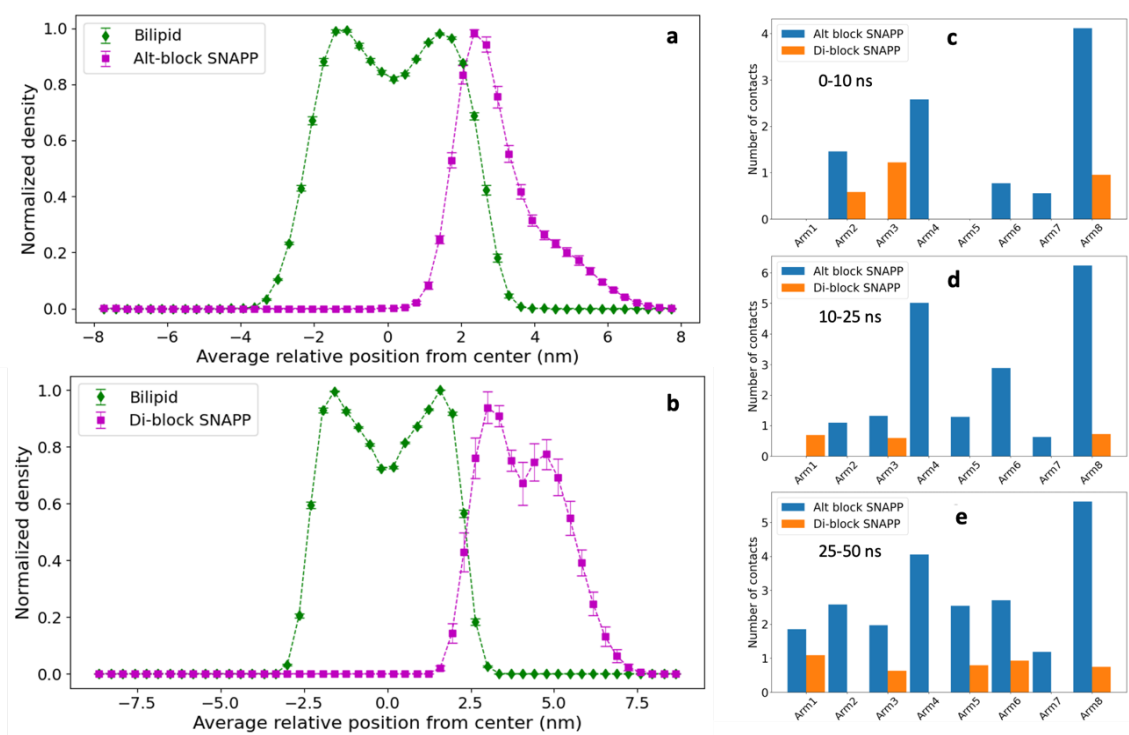

Figure 8: Normalized partial density of SNAPP and bilipid membrane vs. average relative position from the center. (a) Normalized partial density of the alternating block SNAPP and bilipid layer relative to the average position from the center. (b) Normalized partial density of the di-Block SNAPP and bilipid layer relative to the average position from the center. (c) The number of contacts between alternating and di-block SNAPPs with the bilipid membrane are represented in three different distinct time intervals of 0-10 ns, 10-25 ns, and 25-50 ns.

## 4. Conclusion

Through a series of comprehensive atomistic molecular dynamic simulations, we have successfully unravelled the intricate step-by-step mechanism by which alternating-block KKV SNAPP leads to pore formation and disrupts the bilipid structure. We also discuss the relatively lower potency of di-block SNAPP as a membrane disrupting agent. The central finding of our research is that a combination of SNAPP's hydrophobic characteristics and strategic arrangement of hydrophobic and hydrophilic amino acids within its arms are pivotal to its cell disruption mechanism. Furthermore, the spatial arrangement of these amino acids within the arms significantly influences their secondary structure flexibility.

We demonstrated that the randomized placement of hydrophilic and hydrophobic amino acids within SNAPP arms plays a vital role in enhancing its adherence to the bilipid membrane. Specifically, the hydrophilic amino acid lysine exhibits a strong affinity to the lipid head groups within the membrane, forging robust hydrogen bonds. Simultaneously, the hydrophobic amino acid valine interacts with the hydrophobic lipid tails, facilitating SNAPP's integration into the bilipid layer. Notably, this interaction not only involves water molecules near SNAPP but also triggers attraction from water molecules further away within the simulated cellular environment, ultimately leading to formation of pore and destructing the integrity of the bilipid membrane.

The atomistic insights gained from this study hold promise for optimizing the chemical structure of SNAPP. Given the successful validation of our model against existing experimental observations, we are well-positioned to leverage this framework with machine learning and statistical optimization techniques[55,56,57,58,59,60] for designing SNAPP variants with varying numbers of arms and diverse hydrophilic-hydrophobic amino acid sequences. This avenue of exploration could unveil the most effective combinations of amino acids in an arm for SNAPP-mediated disruptions, further advancing our understanding of cellular membrane interactions.

## Acknowledgment

A.J is supported by research training programme (RTP) scholarship provided by the Australian government. This research was supported by The University of Melbourne's Research Computing Services and the Petascale Campus Initiative.

## Conflict of interest

Authors declare no conflict of interest.